\documentclass[twocolumn, floatfix, amsmath, showpacs]{revtex4}
\usepackage{graphicx}

\begin{document}
\title{Focusing of light by random scattering}
\author{I.M. Vellekoop and A. P. Mosk}
\affiliation{Complex Photonic Systems, Faculty of Science and Technology, and MESA$^+$ Research Institute,
University of Twente, P.O.Box 217, 7500 AE Enschede, The Netherlands}

\date{June 21, 2006}

\begin{abstract}
Random scattering of light is what makes materials such as white paint, clouds and biological tissue
opaque. We show that although light propagating in these media is diffuse, a high degree of control is
possible as phase information is not irreversibly lost. Opaque objects such as eggshell or white paint
focus coherent light as sharply as a lens when illuminated with a wavefront that inverts the wave
diffusion. We demonstrate the construction of such wavefronts using feedback, achieving a focus that is
1000 times brighter than the diffusely transmitted light. Our results are explained quantitatively by a
universal relation based on statistical optics.
\end{abstract}\maketitle


Optical microscopy and spectroscopy are essential tools for the study of living organisms and inanimate
objects. These methods rely on the ability to deliver and collect light with a high degree of control.
Unfortunately, in many organic and inorganic materials light is scattered randomly and the required
directionality of the light is lost \cite{Milne1921,Chandrasekhar1960,Ishimaru1978}. In these materials,
light performs a random walk and emerges as a random interference pattern known as speckle. Innovative
imaging methods are directed towards isolating the unscattered fraction of the light
\cite{Vakoc2005,Zhou2006} and towards obtaining useful information from the multiply scattered light
\cite{Li2005,Sebbah2001,Outer1993a}. Ideally, one would completely eliminate or counteract scattering.

We demonstrate inverse wave diffusion, a method for counteracting scattering and diffusion of light. By
constructing a perfectly matched wavefront, we make normally opaque objects focus light as sharply as a
lens. This wavefront cannot be known a priori and is constructed using feedback from a detector in the
target focus. By changing the incoming wavefront, we control the position and shape of the focus; it is
even possible to transmit collimated beams or simple images. Inverse wave diffusion is universally
applicable to scattering objects regardless of their constitution and scattering strength. We envision
that, with such active control, random scattering will become beneficial, rather than detrimental, to
imaging \cite{Lobkis2001}, communication \cite{Derode2003,Simon2001} and non-linear optics
\cite{Baudrier-Raybaut2004}.

\begin{figure}
  \includegraphics[width=8.6cm]{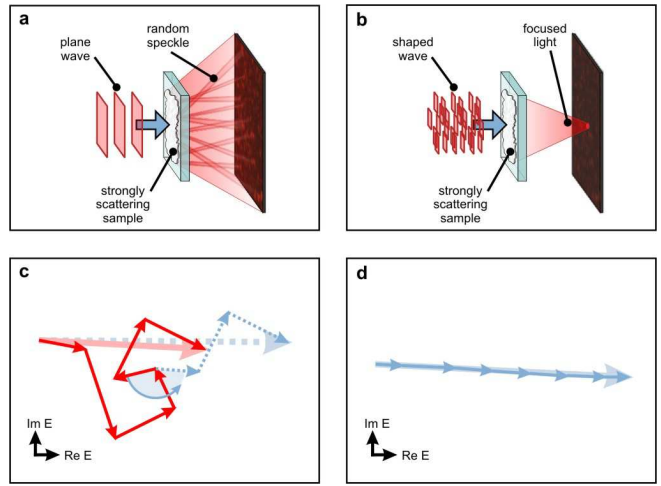}\\
  \caption{Design of the experiment. (a) A plane wave is focused on
  a disordered medium, a speckle pattern is transmitted. (b) The
  wavefront of the incident light is shaped so that scattering
  makes the light focus at any desired point. (c) Complex
  amplitude representation of the field at the target.
  Before optimization, each segment of the incident wavefront
  contributes to the field (thick arrow) in a random way. By
  adjusting the phase of a single segment, we determine the phase
  at which the total field is maximal (dotted arrows). (d) The
  phase of the incident light is adjusted to have all fields
  interfere constructively. The target intensity is at the global maximum.}
  \label{fig:overview}
\end{figure}

Figure \ref{fig:overview} shows the principle of the experiment. Normally, incident light from a 632.8 nm
HeNe laser is scattered by the sample and forms a random speckle pattern (Fig. \ref{fig:overview}(a)). The
goal is to match the incident wavefront to the sample, so that the scattered light is focused in a
specified target area (Fig. \ref{fig:overview}(b)). We divide the incident wavefront into an array of
spatial segments. Initially, the amplitude from each segment contributes randomly to the light amplitude
in the target focus (Fig. \ref{fig:overview}(c)). With a spatial phase modulator, the wavefront of the
incident light is adjusted using the intensity in the focus as feedback. The phase of each segment is
adjusted until the target intensity is maximal. After adjustment, the phase modulator inverts the
diffusion in the sample and the scattered light interferes constructively in the target focus (Fig.
\ref{fig:overview}(d)).

\begin{figure}
  \includegraphics[width=8.6cm]{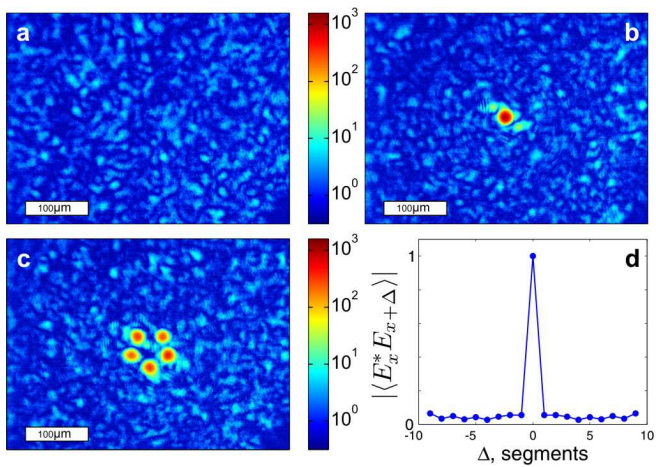}\\
  \caption{Shaped transmission through a strongly scattering sample
  consisting of TiO$_2$ pigment. (a) Transmission micrograph with an unshaped
  incident beam. The scattered light forms a random spackle pattern.
  (b) Transmission after optimization for focusing at a single target.
  The scattered light is focused to a diffraction limited spot that is
  1000 times brighter than the original speckle pattern (c)
  Multi-beam optimization. The disordered medium generates five sharp foci at
  the defined positions. Figures \ref{fig:speckle}(a) to \ref{fig:speckle}(c) are represented on the same logarithmic color
  scale that is normalized to the average transmission before optimization.
  (d) Correlation function of the incident wavefront used to form the pentagon seen in Fig. \ref{fig:speckle}(c).
  There is no correlation between neighboring segments, which indicates that their contributions to the
  target function are independent. The incident wavefronts are composed of a total of 3228 individually controlled segments.} \label{fig:speckle}
\end{figure}

We performed first tests of inverse wave diffusion using rutile TiO$_2$ pigment, which is one of the most
strongly scattering materials known. The sample consists of an opaque, 10.1-$\mu$m thick layer of white
pigment \cite{Kop1997} with a transport mean free path of $0.55 \pm 0.10 \mu$m measured at a wavelength of
632.8 nm (See Appendix A). Since in this sample the transmitted light is scattered hundreds of times,
there is no direct relation between the incident wavefront and the transmitted image
\cite{Pappu2002,Goodman2000}. Figure \ref{fig:speckle} shows the intensity of the transmitted light seen
through a microscope objective. In the first image (Fig. \ref{fig:speckle}(a)) we see the pattern that was
recorded when a plane wave was focused onto the sample. The transmitted light formed a typical random
speckle pattern with a low intensity. We then optimized the wavefront so that the scattered light focused
to a target area with the size of a single speckle. The result is seen in Fig. \ref{fig:speckle}(b), where
a single bright spot stands out clearly against the diffuse background. The focus was over a factor 1000
more intense than the non-optimized speckle pattern. Instead of focusing light to a single spot, it is
also possible to optimize multiple foci simultaneously. By adjusting the target function used as feedback,
the scattered light was made to form simple images, such as the pentagon shown in Fig.
\ref{fig:speckle}(c). Figure \ref{fig:speckle}(d) shows that there is no correlation between neighboring
segments of the optimal incident wavefront, which indicates that the light was fully scattered and no
ballistic transmission occurred.

\begin{figure}
  \includegraphics[width=8.6cm]{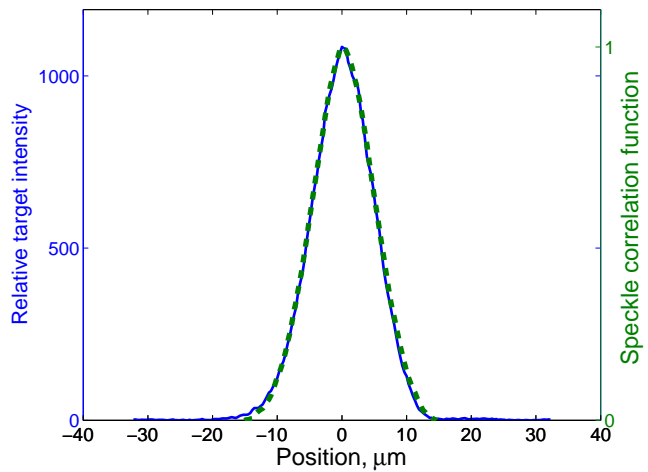}\\
  \caption{Cut through the profile of the target focus after optimization (see Fig. \ref{fig:speckle}(b))
  compared to the normalized autocorrelation function of the speckle before optimization (dotted line).}
  \label{fig:diffraction}
\end{figure}

The TiO$_2$ produced a high quality focus, as can be seen in Figure \ref{fig:diffraction}. For comparison,
the intensity autocorrelation function of the speckle before optimization is also shown. From statistical
optics it is known that the profile of the autocorrelation function equals the diffraction limited beam
profile \cite{Goodman2000}. Since the two functions overlap, we conclude that the multiply scattering
TiO$_2$ produces a diffraction limited focus, i.e. the disordered layer of pigment focuses light as
sharply as an ideal lens of the same size. These results are clear proof that the wavefront of multiply
scattered transmitted light can be controlled with high accuracy.

\begin{table}
    \begin{tabular}{|l|r@{}l|r@{}l|r@{}l|r|r|}
  \hline
    Sample         & \multicolumn{2}{c|}{L ($\mu$m)} & \multicolumn{2}{c|}{d ($\mu$m)} & \multicolumn{2}{c|}{f (mm)}  & \multicolumn{1}{c|}{$\eta\;(\pm15\%)$}   & \multicolumn{1}{c|}{$N$}\\
  \hline
    TiO$_2$        & 10&.1$\pm$0.3   & 18&.7          &3&.5$\pm$0.5    & 1080  & 3228\\
    Petal, fresh   & 43&$\pm$5       & 10&.6          &3&.5$\pm$0.5    &   64  & 208\\
    Petal, dry     & 37&$\pm$5       & 12&.6          &3&.5$\pm$0.5    &  630  & 1664\\
    Eggshell       & 430&$\pm$30     &  2&.1          &3&.5$\pm$0.5    &  250  & 3228\\
    Tooth          & 1500&$\pm$100   & 155&           &125&$\pm$10     &   70  & 208\\
  \hline
    \end{tabular}
    \caption{Intensity enhancement for different materials. $L$, sample thickness ($\pm$ surface roughness), $d$, diameter ($1/e^2$ intensity) of focus,
    $f$, distance between random medium and focus, $\eta$, maximum enhancement reached, $N$, number of segments used by the algorithm to describe the wavefront.}
    \label{tab:enhancements}
\end{table}

Inversion of wave diffusion was realized in a variety of opaque materials ranging from white pigment to
fresh flower petals. Table \ref{tab:enhancements} summarizes the results for the different materials used.
Although the samples vary in thickness, composition and scattering strength, they were all able to focus a
properly prepared wavefront to a diffraction limited spot. The intensity enhancement $\eta$ - defined as
the ratio between the optimized intensity and the average intensity before optimization - varies between
60 and 1000. The main reason for this variation is that the temporal stability of the transmitted speckle
pattern is not the same for all materials.

The optimization procedure makes use of the linearity of the scattering process. The transmitted field in
the target area, $E_m$, is a linear combination of the fields coming from the $N$ different segments of
the modulator,

\begin{equation}
E_m = \sum^N_{n=1}t_{mn}A_n e^{i\phi_n},\label{eq:scattering}
\end{equation}

\noindent where $A_n$ and $\phi_n$ are, respectively, the amplitude and phase of the light reflected from
segment $n$. All scattering in the sample is described by the elements $t_{mn}$ of the unknown
transmission matrix. Clearly, the magnitude of $E_m$ will be the highest when all terms in Eq.
\ref{eq:scattering} are in phase (also see Fig. \ref{fig:overview}(c), \ref{fig:overview}(d)). We
determine the optimal phase for a single segment at a time by cycling its phase from 0 to 2$\pi$. When the
phase of a single segment $n$ is changed, the target intensity detected by the CCD responds as

\begin{align}
|E_m|^2 &=  |E_0|^2  + |t_{mn} A_{n}|^2 +\nonumber\\ &\quad 2 |E_0| |t_{mn} A_n| \cos
\left[\arg(t_{mn})-\arg(E_0)+\phi_n\right],\label{eq:measuring}
\end{align}

\noindent where $E_0$ is the field of the scattered light originating from all segments except segment
$n$. When $N$ is large, each segment contributes little to the total field and $E_0$ is equal for all
segments. For each segment we store the phase at which the target intensity is the highest. At that point
the contribution of segment $n$ is in phase with the already present diffuse background $E_0$. After the
measurements have been performed for all segments, the phase of the segments is set to their stored
values. Now the contributions from all segments interfere constructively and the target intensity is at
the global maximum. This method is generally applicable to linear systems and does not rely on time
reversal symmetry or absence of absorption.

The maximum intensity enhancement that can be reached is related to the number of segments that are used
to describe the incident wavefront. For a disordered medium the constants $t_{mn}$ are statistically
independent and obey a circular Gaussian distribution
\cite{Goodman2000,Garcia1989,Webster2004,Beenakker1997,Pendry1990} and the expected enhancement $\eta$ can
be calculated,

\begin{equation}
\eta = \frac{\pi}{4}(N-1)+1,\label{eq:scaling}
\end{equation}

\noindent It was assumed that all segments of the phase modulator contribute equally to the total incident
intensity. We expect the linear scaling behavior to be universal as Eq. \ref{eq:scaling} contains no
parameters. Neither sample thickness nor scattering parameters will influence the expected intensity
enhancement. Also, since we are free to choose the basis for Eq. \ref{eq:scattering}, we expect to find
the same enhancement regardless of whether the target is a focus or a far-field beam and regardless of how
the shaped wavefront is projected onto the sample; of course the required optimal wavefront will be
different for these varying configurations. The number of degrees of freedom $N$ is bound to a maximum
given by the number independent speckles on the sample surface; $N_\text{max}=8A/\lambda^2$, where $A$ is
the illuminated surface area of the sample and $\lambda$ is the wavelength of the light
\cite{Beenakker1997,Pendry1990}. When $N$ approaches $N_\text{max}$, the intensity in the target focus
alone becomes comparable to the total transmission before optimization. In this extremely interesting
regime, the assumptions underlying Eq. \ref{eq:scaling} are no longer valid. With our current apparatus
$N\ll N_\text{max}$.

\begin{figure}
  \includegraphics[width=8.6cm]{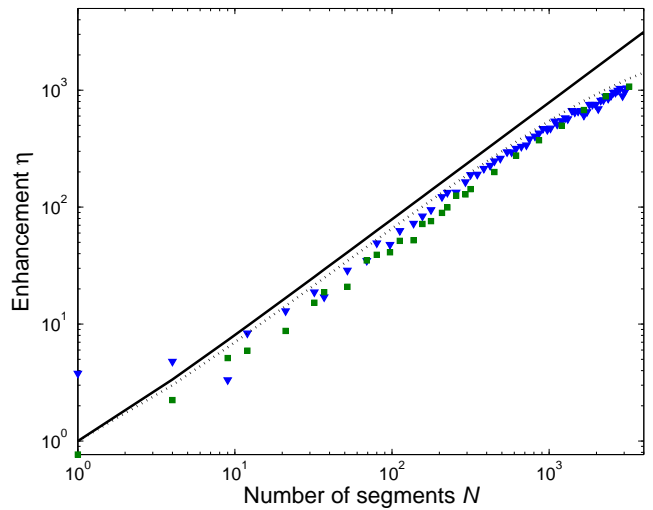}\\
  \caption{Enhancement as a function of the number of
  segments. The experiment was performed twice, once
  with the sample in focus (squares) and once with the sample 100
  $\mu$m out of focus (triangles). The solid line represents the
  theoretical enhancement for an ideal system (Eq. \ref{eq:scaling}). The
  dotted line represents the expected enhancement when residual
  amplitude modulation of the phase modulator and the finite
  persistence time of the sample are taken into account. The uncertainty in $\eta$ is of the order of the symbol size.}
  \label{fig:enhancement}
\end{figure}

We tested the universal scaling behavior implied by Eq. \ref{eq:scaling} by changing the number of
segments into which the phase modulator is subdivided. Using the same TiO$_2$ sample as before, the
algorithm was targeted to construct a collimated beam. In Fig. \ref{fig:enhancement} the enhancement is
plotted as a function of the number of segments for different focusing conditions. The linear relation
between the enhancement and the number of segments is evident until the enhancement saturates at
$\eta=1000$. All measured enhancements were slightly below the theoretical maximum. This is understandable
since the phase space is huge and all perturbations move the system away from the global maximum. The main
reason for deviations from the optimal wavefront is residual amplitude modulation in the phase modulator.
The amplitude modulation introduced a constant, uncontrollable bias in the field (See Appendix A).

The saturation of the enhancement was the result of slow changes in the speckle pattern. This instability
effectively limited the number of segments for which the optimal phase could be measured. We estimate that
the effective enhancement decreases to $\eta_\mathrm{eff} = \eta / (1+NT/T_s)$, where $T=1.2$s is the time
needed for one measurement and $T_s=5400$s is the timescale at which the speckle pattern of the TiO$_2$
sample remains stable. The instability of the speckle pattern was most likely caused by a fluctuating
humidity of the sample \cite{Drobnik1995}. Depending on the environmental conditions, $T_s$ can be
considerably higher and enhancements of over two thousand have been measured overnight.

Our results show that precise control of diffuse light is possible using an optimal, non-iterative
algorithm; light can be directed through opaque objects to form one or multiple foci. The brightness of
the focal spot is explained by a model based on statistical optics. We expect inverse wave diffusion to
have applications in imaging and light delivery in scattering media. Dynamic measurements in biological
tissue are possible when the time required for achieving a focus can be reduced to below 1 ms per segment
\cite{Vakoc2005,Li2005}; we estimate that this timescale is technologically possible with the use of fast
phase modulators. Furthermore, the high degree of control over the scattered light should permit
experimental verification of random matrix theories for the transport of light and quantum particles
\cite{Beenakker1997,Pendry1990}.

We thank Ad Lagendijk for valuable discussions, Willem Vos and Vinod Subramaniam for a critical reading of
the manuscript, and the Photon Scattering group of the Institute for Atomic and Molecular Physics (AMOLF)
for providing samples. This work is part of the research program of the ``Stichting voor Fundamenteel
Onderzoek der Materie (FOM)", which is financially supported by the ``Nederlandse Organisatie voor
Wetenschappelijk Onderzoek (NWO)"

\section{Appendix A: Materials and Methods}
\subsection*{Phase shaping}
The experiments are performed using a polarized 5 mW Helium-Neon laser with a wavelength of 632.8 nm.
Phase modulation is achieved using a Holoeye LCR-2500 twisted nematic liquid crystal reflective spatial
light modulator. The experimental configuration is shown in Figure A1. We illuminate a circular area
containing $3\cdot10^5$ pixels grouped together in square segments. The phase modulator operates in a
phase mostly mode\cite{Davis2002}. In this mode, the field modulation curve can be closely approximated by
a circle in the complex plane. Due to residual amplitude modulation, the centre of this circle is biased
with respect to the origin. The bias corresponds to an uncontrollable fraction of 40\% of the field.
Therefore, the ratio of the controlled intensity to the average total intensity coming from the phase
modulator is $1.0^2$/($0.4^2$ + $1.0^2$) = 0.86. After each adjustment of the phase, we allow for a
stabilization time of 100 ms. The phase shaped beam is spatially filtered to remove higher order
diffraction and focused onto the sample using an objective with a magnification of 63 times and a
numerical aperture of 0.85. Since the solid angle covered by the objective is 0.95$\pi$ and only a single
polarization is used, at most 24\% of the mesoscopic channels can be addressed. A fraction of the phase
shaped light is directed to a silicon photodiode which acts as an intensity reference.

\subsection*{Detection}
The transmitted light is imaged using a microscope objective with an NA of 0.5 and a magnification of 20
times. The light passes a Glan Thompson polarizer and is detected using a 2/3", 12-bit CCD camera in the
back focal plane of the objective. The camera image is integrated over an area that is smaller than the
typical speckle size. No collimating optics are used behind the sample. Per segment of the phase
modulator, the phase is varied between 0 and 2$\pi$ in ten equal steps and a sinusoid is fitted to the
measured intensities. To minimize the effect of noise, the optimization procedure first performs a coarse
pre-optimization using 12 segments. Then, the algorithm is executed twice. The result of the first
iteration is used as the reference field for the second iteration. The average intensity of the speckle
background is obtained by averaging over 4000 random phase patterns.

\subsection*{Samples}
Four different samples were used: TiO$_2$, flower petals, egg shell, and a primary tooth. The first sample
consists of an opaque, 10.1 $\mu$m thick layer of rutile TiO$_2$ pigment on a 2mm-thick fused silica
substrate\cite{Kop1997}. By measuring the total transmission, the transport mean free path was found to be
$0.55 \pm 0.10 \mu$m at a wavelength of 632.8 nm. The sample was placed with the pigment layer towards the
first microscope objective. The flower petal was freshly picked from a {\em{Bellis Perennis}} (Common
Daisy) and fixated with room temperature parafilm between a microscope slide and a coverslip. A second
petal was wet mounted on the slide after which the sample was allowed to dry for one day. The egg shell is
from a white chickens egg. A part of the shell was rinsed, dried and placed between the microscope
objectives. The primary tooth (incisor) was placed in the focal plane of the first microscope objective.
In this latter case, a second objective was not used.

\begin{widetext}
\begin{figure}
  \includegraphics[width=16cm]{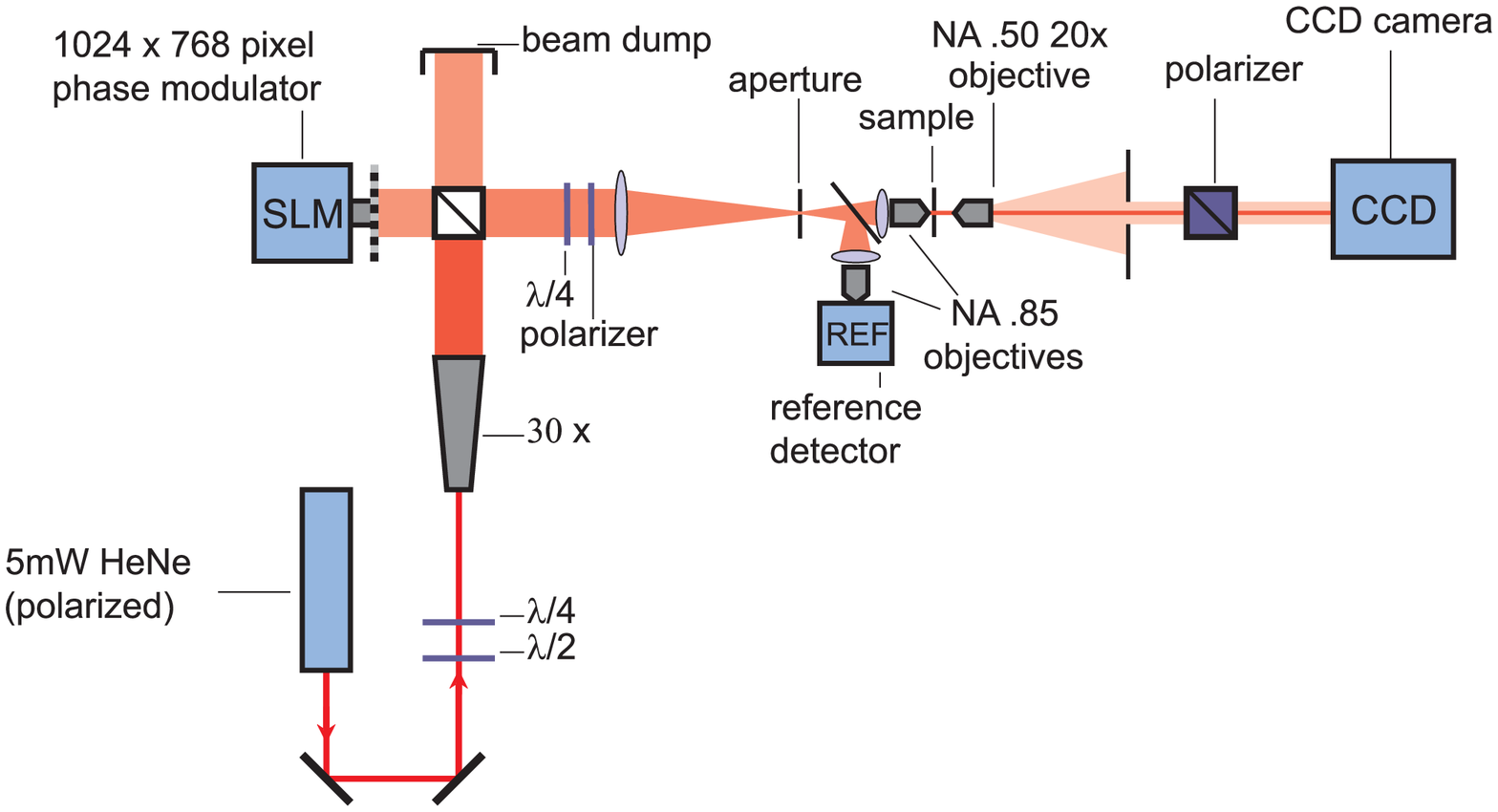}\\
  Figure A1: Schematic drawing of the apparatus.
  \label{fig:schematic}
\end{figure}
\end{widetext}

\bibliography{../../bibliography}
\bibliographystyle{apsrev}

\end{document}